\documentclass[conference]{IEEEtran}
\IEEEoverridecommandlockouts
\usepackage{cite}
\usepackage{amsmath,amssymb,amsfonts}
\usepackage{algorithmic}
\usepackage{graphicx}
\usepackage{textcomp}
\usepackage{xcolor}
\usepackage{float}

\usepackage[font=small,skip=0pt]{caption}
\def\BibTeX{{\rm B\kern-.05em{\sc i\kern-.025em b}\kern-.08em
    T\kern-.1667em\lower.7ex\hbox{E}\kern-.125emX}}
\begin{document}
\title{Investigating Application of Deep Neural Networks 
\newline in Intrusion Detection System Design}

\author{\IEEEauthorblockN{Mofe O. Jeje}
\IEEEauthorblockA{\textit{Department of Computer Science} \\
\textit{School of Electrical Engineering and Computer Science} \\
\textit{University of North Dakota}\\
Grand Folks, North Dakoka, United States \\
mofe.jeje@ndus.edu}}


\maketitle

\begin{abstract}
     Despite decades of development, existing IDSs still face challenges in improving detection accuracy, evasion, and detection of unknown attacks. To solve these problems, many researchers have focused on designing and developing IDSs that use Deep Neural Networks (DNN) which provides advanced methods of threat investigation and detection. Given this reason, the motivation of this research then, is to learn how effective applications of Deep Neural Networks (DNN) can accurately detect and identify malicious network intrusion, while advancing the frontiers of their optimal potential use in network intrusion detection. Using the ASNM-TUN dataset, the study used a Multilayer Perceptron modeling approach in Deep Neural Network to identify network intrusions, in addition to distinguishing them in terms of legitimate network traffic, direct network attacks, and obfuscated network attacks. To further enhance the speed and efficiency of this DNN solution, a thorough feature selection technique called Forward Feature Selection (FFS), which resulted in a significant reduction in the feature subset, was implemented. Using the Multilayer Perceptron model, test results demonstrate no support for the model to accurately and correctly distinguish the classification of network intrusion.
\end{abstract}

\section{INTRODUCTION}
    Given the ever-growing complication and severity of security attacks on computer networks to which traditional intrusion detection system is vulnerable, there has been a considerable interest in the task of designing and constructing comprehensive IDS framework solution, that can adequately defend and protect organizations’ data and information asset.
    
\medskip While this interest, inspired security researchers to propose various detection paradigms that incorporate different machine learning methods, in terms of classification techniques, statistical theories, and information theories, among others; the interplay of a quickly changing network environments, having attack variants with novel attacks emerging constantly, together with a variety of network types which generate large-scale data with high-dimensional structures, further compounded an already complex issue, for which traditional intrusion detection approaches are unsuitable [1].

\medskip Despite decades of incorporating machine learning methods, existing intrusion detection system (IDS) still face challenges in improving the detection accuracy, error classification, and detection of unknown attacks, - all of which are birthed by the difficulty of designing representative traffic features, that are in turn occasioned by the lack of standard guiding principle for the design of a feature set that can accurately characterize network traffic [2].

\medskip However, while the design of a feature set that can accurately characterize network traffic is still an ongoing research issue [3], many researchers have focused on developing IDS that leverage on Deep learning which has the potential of dealing with the constantly dynamic and ever-evolving network threats, as the model, compared to other shallow models of machine learning, is touted to have stronger fitting ability, and can improve IDS performance in many aspects. 

\medskip Owing to it being a promising method for solving the problems inherent with anomaly intrusion detection system, in terms of detecting unknown and new attacks; modeling with deep structures for classification, feature extraction, feature reduction, data denoising, and data augmentation tasks; has attracted increasing interest, in addition to becoming an outstanding direction of study in academic, government and industrial circles.

\medskip The above revelation, and the decent results achieved recently with Convolutional Neural Network (CNN) through applications in Natural Language Processing (NLP) and Image Recognition, has essentially inspired an investigation into this research area. The rest of the paper is organized as follow: section 2 reviews related works in the literature; section 3 discusses the research methodology; section 4 covers evaluation and discussion of experimental results; finally, section 5 draws the conclusion and the research future direction. 

\section{RELATED WORKS}
\medskip Previous methods of traffic classification [4], are not competent enough for modern traffic environment due to their disability towards encrypted traffic. The newest generation of traffic classification method is based on flow-statistics and Machine Learning (ML), which can cope with both encrypted and normal traffic.

\medskip However, as the performance of ML-Based approaches highly depends on the human-engineered features, and some private traffic information, their use dramatically suffers from accuracy and generalizability limitations. 

\medskip To overcome this issue of accuracy and strength of generalizability, the application of deep learning is touted to offer a better outcome, achieving a great deal of attention in the IDS domain, with numerous deep learning-based misuse detection and anomaly detection models being provided in the literature to deal with various types of intrusions and security attacks. This possibility has led the way for more successful application of deep learning in various domains such as text, audio, and visual processing [5], as well as other contexts, such as sentiment analysis [6], social network analysis, recommender systems, natural language processing, and wireless networking [7], among others.

\medskip Its preference by many IDS/IPS researchers was born out of its higher learning capability in comparison with traditional ML methods like Random Forest, Support Vector Machine, and KNN to mention but a few [8]. In other words, this means that deep learning possesses the potential to learn highly complicated patterns to gain a higher accuracy than previous methods with more functionality. 

\medskip While early research with the application of Convolutional Neural Network in designing a multiple classification IDS failed to take full advantage of deep neural networks, resulting from the constraint that these early research mainly focussed on models that only learned features from a manually designed traffic features without raw network data; recent applications in the fields of computer vision and natural language processing have shown that the biggest advantage of deep neural networks lies in their ability to learn features directly from raw data [9]. This way, methods based on Deep Learning (DL) offer advantage that obviate the burdensome work of selecting features and acquiring private features information, as it automatically extracts and selects features through training [10]. This characteristic, no doubt, has made DL-based methods a highly desirable approach for traffic classification. 

\medskip In Corroborating this further, [11] argues that, while feature engineering depends on domain knowledge, and the fact that quality of features often becomes a bottleneck of detection effects; Deep learning-based detection methods can learn feature automatically, and can directly process raw data, that allows them to learn features and perform classification at the same time. These types of methods according to [11] work well in an end-to-end fashion, and are gradually becoming the mainstream approach in IDS studies.
 
\medskip As a caution, while it is of note that Deep learning models have made great strides in big data analysis, their performances are however, according to [12] not ideal on small or unbalanced datasets. Instead, only Adversarial learning approaches can improve the detection accuracy in the context of a small datasets. 

\medskip In sum, the interplay of all the above research studies, particularly as it relates to the use of raw input features dataset, in addition to the automatic extraction and selection of features through training; benefited this research work, and by extension, motivated it to investigate and use deep neural networks to learn features of raw data of network traffic to perform an intrusion detection task.

\section{METHODOLOGY}
\subsection{Model}
The model used in this research uses Multi-Layer Perceptron, a type of feedforward deep neural network that consists of at least three layers of nodes, in terms of an input layer, one or more hidden layers, and an output layer. As shown in figure 1 below, each node, or neuron, in one layer connects with a certain weight to every node in the following layer, making the network fully connected. This research, therefore leverages on the model’s supervised learning technique called backpropagation for its training.

\begin{figure}[H]
\centerline{\includegraphics{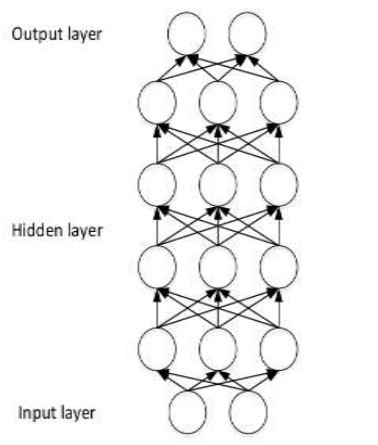}}
\renewcommand{\thefigure}{1}%
\caption{The structure of a Deep Neural Network}
\label{fig}
\end{figure} 

\subsection{Data Characteristics}
Using the ASNM-TUN dataset provided by Brno University Security Laboratory Research Group (BUSLAB), the study thereon uses the Multilayer Perceptron (MLP) to develop a detection model that is capable of accurately detecting network intrusion instances. 

\medskip In description, the dataset consists of ASNM features extracted from tcpdump capture of obfuscated malicious and legitimate TCP communications on selected vulnerable network services. The ASNM-TUN dataset contains 892 features with 4 types of labels totaling 394 rows × 896 columns. 

\medskip But for the purpose of this research study, only 2 out of the 4 labels were used, namely the two-class label, and the three-class label. Also, out of the 892 features, only 20 features were extracted and used for this research purpose.

\subsection{Data Preprocessing Procedure}
As the model requires only numerical inputs, label encoding was carried out only on the Label2 feature which is the only feature requiring conversion from categorical data to numerical data. Also, before using StandardScaler technique to normalize and transform the features to be on a similar scale, with an intent of improving performance, as well as training stability of the model; the dataset was split into 80 percent and 20 percent for training and testing respectively with the use of a train-test-split function of sklearn module in python.

\medskip After data normalization, while an MLP classifier was trained with tuned hyperparameters using the training dataset, the classifier on the other hand was tested using the testing dataset to evaluate the model accuracy, along with a confusion matrix to show its performance.

\medskip To further improve on the model, a Forward Feature Selection (FFS) technique was used to select an optimal subset of features from all the feature candidates that were previously used. The resulting optimal subset of features were subsequently split into 80 percent training set and 20 percent testing set. Following after, the MLP classifier was again trained and evaluated respectively on the resulting optimal subsets, in addition to combining the use of confusion matrix to visualize the model's classificationand detection performance.

\section{DISCUSSION OF EXPERIMENTAL RESULTS} 
\vspace{-12pt}
In discussing evaluation of results, the study begins this section with a confusion matrix. In the tables given below, while the columns are the predictions, and the rows are the actual values; the main diagonal however gives the correct predictions, as it shows cases where the actual values and the model predictions are the same.
\begin{table}[H]
\renewcommand{\thetable}{1a}%
\caption{Confusion matrix for label2}
\begin{center}
\begin{tabular}{|c|c|c|}
\hline
\textbf{Actual}&\multicolumn{2}{|c|}{\textbf{Predicted}} \\
\hline 
& \textbf{\textit{Direct Attack}}& \textbf{\textit{Legitimate Traffic}} \\
\hline
\textbf Direct Attack& 25& 15 \\
\textbf Legitimate Traffic& 3 & 36 \\
\hline
\end{tabular}
\label{table:1}
\end{center}
\end{table}

\vspace{-15pt}
In table 1(a), the first rows are the actual ‘Direct Attacks.’ As seen from the table, the model predicted and classified 25 of these correctly and incorrectly predicted and classified 15 of the ‘Direct Attacks’ as ‘Legitimate Traffics.’ Looking at the ‘Direct Attacks’ column; of the 28 ‘Direct Attacks’ predicted and classified by the model (i.e. the sum of the ‘Direct Attacks’ column), 25 were correctly predicted and classified as ‘Direct Attacks’, while 3, instead of being reported as ‘Legitimate Traffics’ were incorrectly predicted and classified as ‘Direct Attacks’, and likewise in the ‘Legitimate Traffic’ column, 36 were correctly predicted and classified as ‘Legitimate Traffics’ while 15, instead of being reported as ‘Legitimate Traffic’ were incorrectly predicted and classified as ‘Direct Attacks’. 
\begin{table}[H]
\renewcommand{\thetable}{1b}%
\caption{Confusion matrix for label3}
\begin{center}
\begin{tabular}{|c|c|c|c|c|}
\hline
\textbf{Actual}&\multicolumn{3}{|c|}{\textbf{Predicted}} \\
\hline
& \textbf{\textit{Direct Attack}}& \textbf{\textit{Legitimate Traffic}}& \textbf{\textit{Obfuscated Attack}} \\
\hline
\textbf Direct& 11& 0& 9 \\
\textbf Legitimate& 1& 1& 17 \\
\textbf Obfuscated& 0& 0& 40 \\
\hline
\end{tabular}
\label{table:1}
\end{center}
\end{table} 

\vspace{-30pt}

\medskip In table 1(b), just as stated in table 1(a), the columns are the predictions and the rows are the actual values. The main diagonal (11, 1, 40) gives the correct predictions and classifications. In other words, they are the cases where the actual values and the model predictions are the same.

\medskip From the said table, the first rows are the actual ‘Direct Attacks.’ Here, the model predicted and classified 11 of these correctly, but incorrectly predicted and classified nothing or 0 of the ‘Direct Attacks’ as ‘Legitimate Traffics,’ as well as 9 of the ‘Direct Attacks’ misrepresented as ‘Obfuscated Attacks’. Looking at the ‘Direct Attacks’ column; of the 12 ‘Direct Attacks’ predicted and classified by the model (i.e. 
the sum of the ‘Direct Attacks’ column), 11 were correctly predicted and classified as ‘Direct Attacks’, while 1, instead of being reported as ‘Legitimate Traffics’ was incorrectly predicted and classified as ‘Direct Attacks’, and nothing or 0, instead of being reported as ‘Obfuscated Attacks’ was incorrectly predicted and classified as ‘Direct Attacks’. Analogous interpretations apply to the rest of other columns and rows.
\vspace{4pt}

\begin{table}[htbp]
\renewcommand{\thetable}{2a}%
\caption{Classification Report for label2}
\begin{center}
\begin{tabular}{|c|c|c|c|c|c|}
\hline
& \textbf{\textit{Precision}}& \textbf{\textit{Recall}}& \textbf{\textit{F1Score}}& \textbf{\textit{Support}} \\
\hline
\textbf 0& 0.89& 0.62& 0.74& 40 \\
\textbf 1& 0.71& 0.92& 0.80& 39 \\
&&&&\\
\textbf accuracy& & & 0.77& 79 \\
\textbf macro avg& 0.80& 0.77& 0.77& 79 \\
\textbf weighted avg& 0.80& 0.77& 0.77& 79 \\
\hline
\end{tabular}
\label{table:2}
\end{center}
\end{table}

\vspace{-30pt}
\medskip In the classification report of table 2(a) given above, while the study under Label2 reported an accuracy score of 0.77 along with an F1score of 0.80, it however, achieved a precision score of 0.71 and a recall score of 0.92.

\vspace{3pt}

\begin{table}[H]
\renewcommand{\thetable}{2b}%
\caption{Classification Report for label3}
\begin{center}
\begin{tabular}{|c|c|c|c|c|c|}
\hline
& \textbf{\textit{Precision}}& \textbf{\textit{Recall}}& \textbf{\textit{F1Score}}& \textbf{\textit{Support}} \\
\hline
\textbf 1& 0.92& 0.55& 0.69& 20 \\
\textbf 2& 1.00& 0.05& 0.10& 19 \\
\textbf 3& 0.61& 1.00& 0.75& 40 \\
&&&&\\
\textbf accuracy& & & 0.66& 79 \\
\textbf macro avg& 0.64& 0.53& 0.51& 79 \\
\textbf weighted avg& 0.78& 0.66& 0.58& 79 \\
\hline
\end{tabular}
\label{table:2}
\end{center}
\end{table}

\vspace{-30pt}
\medskip Also in the classification report of table 2(b) given above, while the study under Label3 reported an accuracy score of 0.66 along with a weighted average of F1score of 0.58, it reported a precision weighted average of 0.78 and a recall weighted average of 0.66.
\vspace{0.00mm}
\begin{figure}[H]
\centerline{\includegraphics{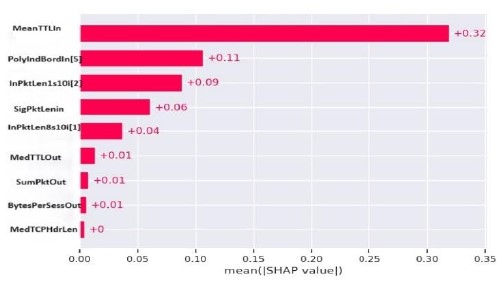}}
\renewcommand{\thefigure}{2a}%
\caption{Summary Plot of Feature Importance Label2}
\label{fig}
\end{figure}

\vspace{-12pt}
\begin{figure}[H]
\centerline{\includegraphics{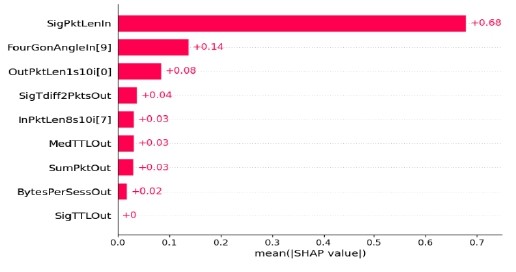}}
\renewcommand{\thefigure}{2b}%
\caption{Summary Plot of Feature Importance Label3}
\label{fig}
\end{figure} 

\setlength{\intextsep}{-9pt}

\medskip As shown above in figure 2(a) of Label2, the Mean absolute SHAP values are typically displayed as bar plots that rank features by their importance. The main characteristics to examine here are the ordering of features, and the relative magnitudes of the mean absolute SHAP values. Here, MeanTTLIn is seen as the most influential variable, contributing on average ±0.32 to each predicted and classified label2 observations, in terms of Direct Attacks, and Legitimate traffics. By contrast, the least influential variable is MedTCPHdrLen which contributes ±0 or nothing to the observations.\vspace{10pt}

\medskip In figure 2(b) of Label3 however, the inclusion of Obfuscated Attacks along with Direct Attacks, as well as Legitimate traffics, see SigPktLenin as the most influential variable, contributing on average ±0.68 to each predicted and classified observation under label3. On the other spectrum of the plot, the least influential variable is SigTTLOut which contributes ±0 or nothing to the observations.

\vspace{15pt}

\begin{figure}[H]
\centerline{\includegraphics{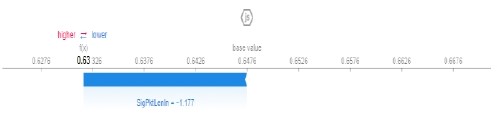}}
\renewcommand{\thefigure}{3a}%
\caption{Force Plot (Label2)}
\label{fig}
\end{figure}

\vspace{15pt}
\begin{figure} [H]
\centerline{\includegraphics{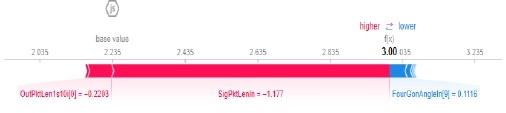}}
\renewcommand{\thefigure}{3b}%
\caption{Force Plot (Label3)}
\label{fig}
\end{figure}

\medskip In the above figures, each stripe shows the impact of its feature in pushing the value of the target variable farther or closer to the base value. While Red stripes show that their features push the value towards higher values, the Blue stripes on the other hand, show that their features push the value towards lower values. So, the wider a stripe, the higher its contribution. Therefore, the sum of the contributions from all stripes pushes the value of target variable from the base value to the final predicted value. 

\vspace{-45pt}
\medskip In the force plot for Label2 in figure 3(a) involving only Direct Attacks and Legitimate traffics, only SigPktLenin is reported as an observation with a long Blue strip showing negative contribution to the predicted value of 0.63 and a base value of 0.645. 
\vspace{-40pt}
\setlength{\intextsep}{68pt}

\medskip However, in the force plot for Label3 in figure 3(b) involving Direct Attacks and Legitimate traffics, along with the inclusion of Obsfucated Attacks; we see, OutPktLen1s10i[0], and SigPktLenin values having a positive contribution to the predicted value of 3.00 and a base value 2.234, which is the average value of the target variable across all the records. Out of the 2 variables, SigPktLenin is the most important variable of record, showing positive contribution with the widest stripe (it has the largest stripe). Even though FourGonAnglein[9] is the only variable that shows a negative contribution, the value of its contribution is not strong enough to move the predicted value lower than the base value. So, since the total positive contribution (Red stripes) is larger than the negative contribution (Blue stripe), the final value is greater than the base value.

\setlength{\textfloatsep}{-50pt}

\vspace{-30pt}
\begin{figure}[H]
\centerline{\includegraphics{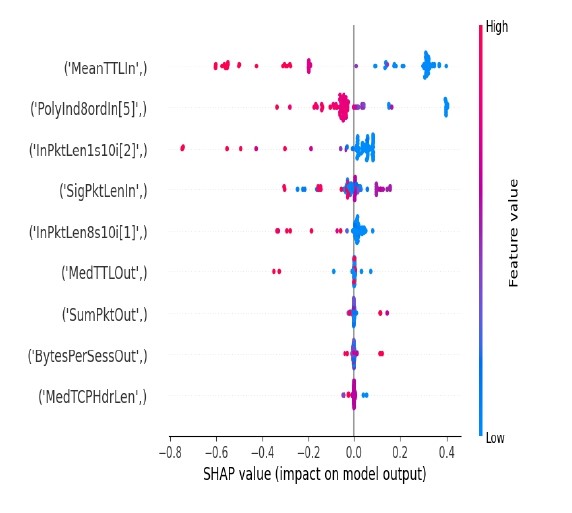}}
\renewcommand{\thefigure}{4a}%
\caption{Beeswarm Plots (Label2)}
\label{fig}
\end{figure}

\setlength{\belowcaptionskip}{2ex} 
\vspace{-20pt}
\begin{figure}[H]
\centerline{\includegraphics{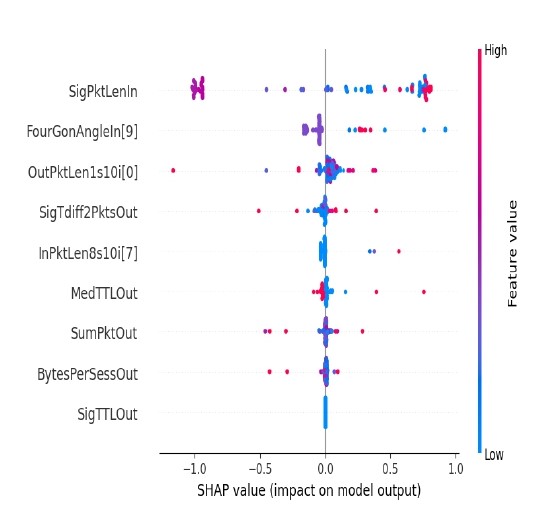}}
\renewcommand{\thefigure}{4b}%
\caption{Beeswarm Plots (Label3)}
\label{fig}
\end{figure}

\setlength{\intextsep}{46pt}
\vspace{-130pt}

\medskip From the two Beeswarm figures above, the features are sorted from the most important one to the least important. The higher the value of a given feature, the more positive the impact on the target. The lower this value, the more negative the contribution. 

\vspace{-50pt}
\medskip In figure 4a of Label2 therefore, while MeanTTLIn can be seen as the most important feature, followed by PolyIn8ordIn[5] and so on; MedTCPHdrLen is shown to be the least important feature. Figure 4b of Label3, in contrast has SigPktLeanin as the most important feature, followed by FourGonAnglein[9] and so on, while the plot on the other side of the spectrum, reveals SigTTLOut as the least important feature. 

\vspace{-15pt}
\begin{figure}[H]
\centerline{\includegraphics{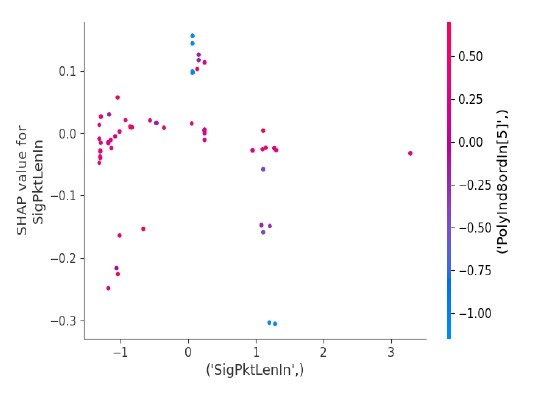}}
\renewcommand{\thefigure}{5a}%
\caption{Dependence Plot for SigPktLenin (label2)}
\label{fig}
\medskip\end{figure}

\begin{figure}[H]
\centerline{\includegraphics{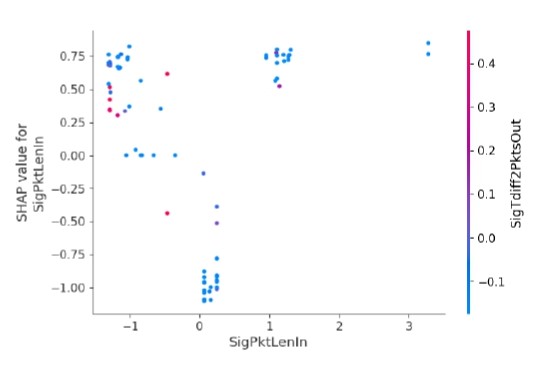}}
\renewcommand{\thefigure}{5b}%
\caption{Dependence Plot for SigPktLenin (label3)}
\label{fig}
\end{figure}


\setlength{\intextsep}{130pt}
\vspace{-70pt}

\medskip While in the dependence plot of figure 5a of Label2, a network feature, named SigPktLenin can be seen to show a mostly positive effect on the model, along with a higher indication or likelihood of being classified as an attack; figure 5b of Label3 on the other hand reveals the same network feature, named SigPktLenIn showing a mostly negative impact on the model, along with a lower indication or likelihood of not being classified as an attack.

\vspace{-20pt}
\setlength{\intextsep}{20pt}

\setlength{\belowcaptionskip}{-10pt}
\begin{figure}[htbp]
\centerline{\includegraphics[width=1\linewidth]{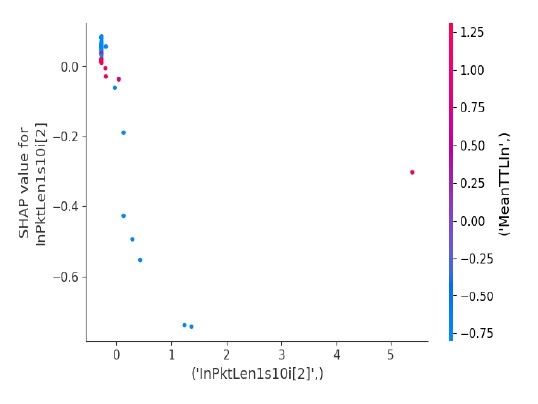}}
\renewcommand{\thefigure}{6a}%
\caption{Dependence Plot for InPktLen8s10i[7] (Label2)}
\label{fig}
\end{figure}

\vspace{-50pt}
\setlength{\intextsep}{40pt}

\begin{figure}[htbp]
\centerline{\includegraphics[width=1\linewidth]{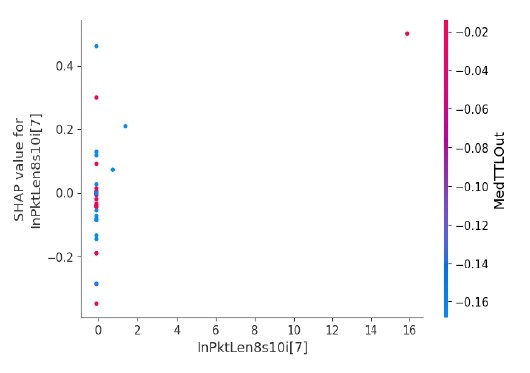}}
\renewcommand{\thefigure}{6b}%
\caption{Dependence Plot for InPktLen8s10i[7] (Label3)}
\label{fig}
\end{figure} 

\setlength{\belowcaptionskip}{10pt}

\setlength{\intextsep}{80pt}
\medskip In figure 6a of Label2, an examination of the dependence plot for InPktLen8s10i[7] reveals a wider presence of Blue which indicates a lower likelihood of network traffic being acknowledged or classified as attacks. With this same variable, an examination of figure 6b of Label3 reveals a scenario of mixed classification wherein some network traffics are acknowledged as attacks, and some also are ignored as attacks.

\setlength{\intextsep}{-2pt}
\begin{figure}[htbp]
\centerline{\includegraphics{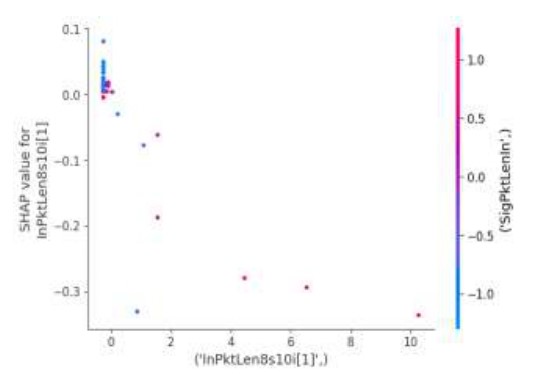}}
\renewcommand{\thefigure}{7a}%
\caption{Dependence Plot for OutPktLen1s10i[0] (Label2)}
\label{fig}
\end{figure}

\begin{figure}[htbp]
\centerline{\includegraphics{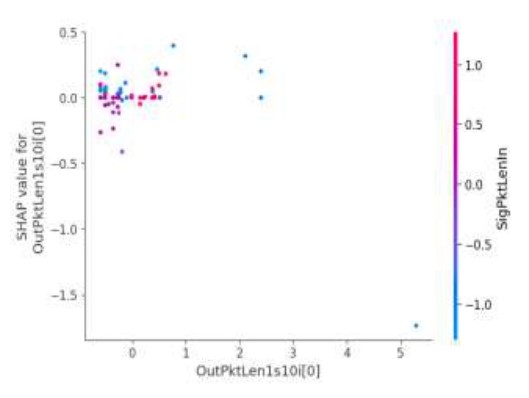}}
\renewcommand{\thefigure}{7b}%
\caption{Dependence Plot for OutPktLen1s10i[0] (Label3)}
\label{fig}
\end{figure}

\vspace{-15pt}

\medskip For figure 7a of Label2, the dependence plot reveals a network feature named InPktLen8s10i[1] with values above zero, generally showing a likelihood of being classified as Attacks than the one below zero. However, a different scenario in figure 7b of Label3 is seen with a network feature named OutPktLen1s10i[0] which generally shows no clarity between the likelihood of being classified as Attacks, and the ones being ignored as Attacks.

\medskip While the examination of dependence plot in figure 8a of Label2 reveals a network feature named PolyIn8ordIn[5] with a general widespread of Blue values showing a lower likelihood of being acknowledged or classified as an Attack; a different picture is seen in figure 8b of Label3. Under Zero, the dependence plot in figure 8b reveals a network feature named FourGonAngleIn[9] having a higher likelihood of Attack classification; but while greater than zero, the dependence plot shows a lower likelihood of an attack classification.

\setlength{\intextsep}{2pt}

\begin{figure}[htbp]
\centerline{\includegraphics{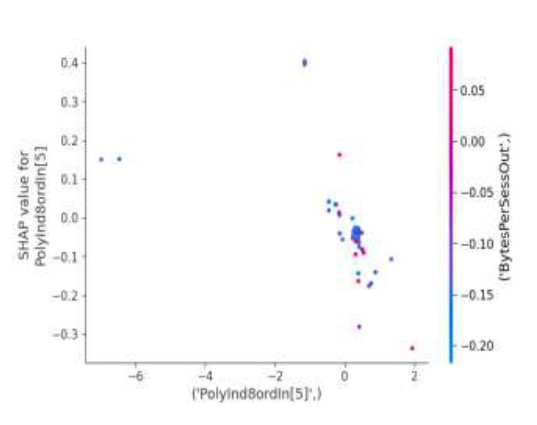}}
\renewcommand{\thefigure}{8a}%
\caption{Dependence Plot for FourGonAngleIn[9]  (Label2}
\label{fig}
\end{figure}

\setlength{\intextsep}{2pt}

\begin{figure}[htbp]
\centerline{\includegraphics{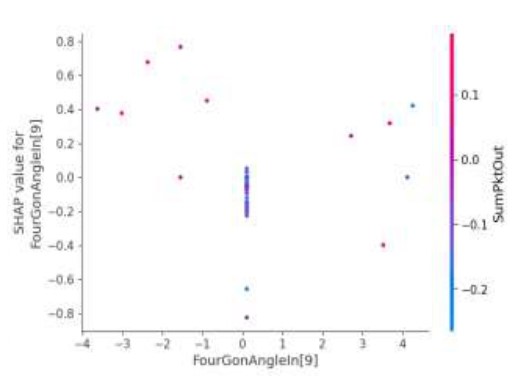}}
\renewcommand{\thefigure}{8b}%
\caption{Dependence Plot for FourGonAngleIn[9]  (Label3}
\label{fig}
\end{figure}

\setlength{\intextsep}{-100pt}
\vspace{-2pt}

\medskip In the Heatmap figures below, each square shows the correlation between the features on each axis with correlation ranging from -1 to +1. Values closer to zero means there is no linear relationship between the two features. The close to 1 the correlation is, the more positively correlated they are; meaning that, as one feature increases, so does the other; and the closer to 1 their correlation value, the stronger their relationship. Same with a correlation value closer to -1, but instead of both features increasing, one feature will decrease as the other increases. 

\vspace{4pt}

\medskip From the heatmap in figure 9a of Label2, while the diagonals are all 1 or dark brown, denoting a perfect correlation, as the squares are correlating each feature to itself; BytesPerSessOut and SumPktOut are seen as the most positively correlated input features with a correlation value of 0.5, and the least negatively correlated input features are MedTTLOut and MedTCPHdrLen with a correlation value of -0.7. For the rest, the larger the number, and the darker the color; the higher the correlation between the two features. For all intents and purposes, the plot is generally symmetrical about the diagonal, since the same two features are being paired together in those squares.

\setlength{\intextsep}{3pt}

\begin{figure}[htbp]
\centerline{\includegraphics{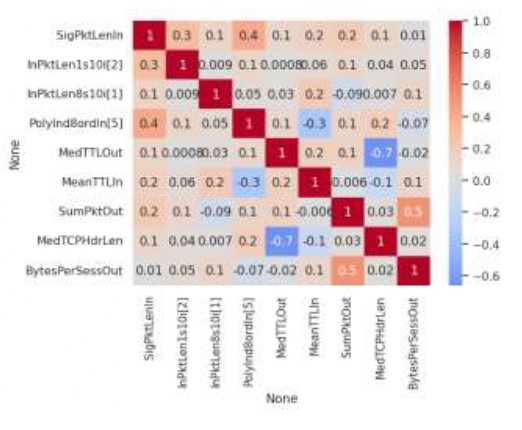}}
\renewcommand{\thefigure}{9a}%
\caption{Heatmap for (Label2)}
\label{fig}
\end{figure}

\setlength{\intextsep}{-5pt}

\begin{figure}[htbp]
\centerline{\includegraphics{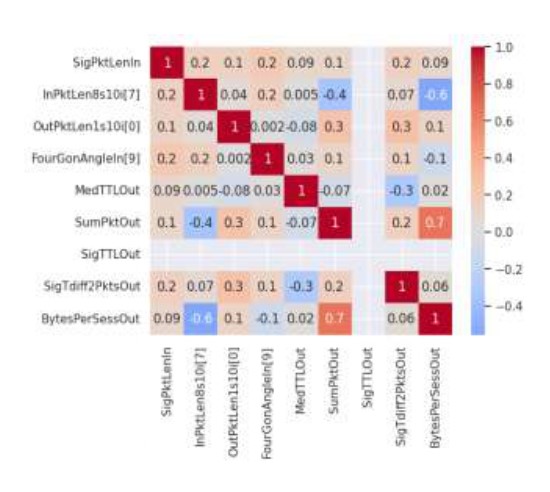}}
\renewcommand{\thefigure}{9b}%
\caption{Heatmap for (Label3)}
\label{fig}
\end{figure}

\medskip In figure 9b of Label3 however, while the left diagonal has all 1 or dark brown, denoting a perfect correlation, as the squares are correlating each feature to itself; SumPktOut and BytesPerSessOut are seen as the most positively correlated input features with a correlation value of 0.7, and the least negatively correlated features are BytesPerSessOut and InPktLen8s10i[7] with a correlation value of -0.6. For the rest, the larger the number, and the darker the color; the higher the correlation between the two features. Apart from SigTTLOut which has no relationship with any features under consideration, as it is blank about its horizontal and vertical axes; the plot is generally symmetrical about the diagonal, since the same two features are being paired together in those squares.
 
\subsection*{Findings and Insights}
\medskip The experiment in the study consisted of two prongs of Forward Feature Selection (FFS) executions. The first execution took as input features only Legitimate Traffic and Direct Attack entries, which represents the case where MLP was trained without knowledge about Obfuscated Attacks (it is called label2). The second execution took as input feature, the whole dataset of network traffic (called label3) which represents the case where MLP was aware of Obfuscated Attacks, along with Legitimate Traffics, as well as Direct Attacks. 

\medskip Before the application of Forward Feature Selection (FFS) technique, findings with label3 show the model with an accuracy score of 0.92 along with a Confusion matrix of 19, 17, and 37 in the right diagonal and 1, 17, and 1 in the left diagonal. Findings with label2 before the application of FFS on the other hand, revealed an accuracy score of 0.92 along with a confusion matrix of 34, and 39 on the right diagonal and the value of 6, and 0 on the left diagonal.

\medskip But after the application of FFS however, findings under Label2, in terms of precision reveal that, out of all the network traffics that final model of the FFS case predicted would be classified as Direct Attacks, only 0.71 could be classified. Also, in terms of recall for all the network traffic that did get classified as Direct Attacks, the model, only predicted this outcome correctly for 0.92 of the time. Also, while an accuracy score of 77 percent was reported, an F1score of 0.8 was achieved, indicating that, the model whose value tends to, or moves close to 1, did a good job in predicting the likelihood of network traffic being correctly classified as Attacks.

\medskip Similarly, findings, in terms of precision under label3 revealed that, out of all the network traffics that the final FFS model predicted, only 0.92 did get correctly classified as Direct Attacks, 1.00 were correctly classified as Obfuscated Attack, and only 0.61 were correctly classified as Legitimate traffic. The model further, in terms of recall, only predicted 0.55 of network traffic correctly as Direct Attacks, 0.05 of network traffic as Obfuscate Attacks, and 1.00 of network traffics as Legitimate Traffic. Furthermore, while an accuracy score of 66 percent was generally reported, an F1score of 0.7, 0.1, and 0.8 were respectively observed under Direct Attacks, Obfuscate Attacks, and Legitimate Traffic, with the model indicating that, the MLP, except for the Obfuscate Attacks, did a good job of predicting the likelihood of network traffics being correctly classified. For the Obfuscated Attacks, the reported values of 0.1 shows that, training the classifier with the inclusion of Obfuscated Attacks, unlike the case of their non-inclusion has been comparably difficult. For this reason, several similar input features were selected in both cases, particularly given the fact that they still offer value regardless of whether obfuscation was performed or not.

\section{CONCLUSION}
\medskip Given the fact that the model did not perform well in terms of accuracy, as its performance value poorly oscillates between 66 percent and 73 percent, in addition to the model being a weak means of detecting Obfuscated Attacks; it then can be safely said, that the model, on one hand, should be used with caution; and on the other hand, it should be optimally used in combination with other models to achieve good performance.

\medskip Additionally, it should be bored in mind that Machine learning methods have a certain ability to detect intrusions, but they often do not perform well on completely unfamiliar dataset. Consequently, when the dataset does not cover all typical real-world samples, good performance in actual environments is not guaranteed, even if the models achieve high accuracy on the test sets.

\subsection*{Future Research Direction}
\medskip The first take away in terms of future research direction for this study is the daunting challenge of runtime, time constraint and storage limitation. These issues underscore the need for practical IDSs to, not only have high detection accuracy, but also need to have high runtime efficiency, because in attack detection, the real-time requirement is of utmost important. 

\medskip Following from this therefore, further research is needed to improve the efficiency of Multi-Layer Perceptron (MLP) with increased storage and computational power.

\medskip Lastly, another research direction could also be to investigate effectiveness, and detection capabilities of the Multi-Layer Perceptron (MLP) with increased input features above 20, along with the multiplicity of Deep Learning models for comparative analysis.

\end{document}